\documentclass[aps,prd,twocolumn,floatfix,superscriptaddress]{revtex4}
\usepackage{graphicx,amssymb,url}

\newcommand{\beq}{\begin{equation}}
\newcommand{\eeq}{\end{equation}}
\newcommand{\ba}{\begin{eqnarray}}
\newcommand{\ea}{\end{eqnarray}}

\def\lsim{\raise0.3ex\hbox{$\;<$\kern-0.75em\raise-1.1ex\hbox{$\sim\;$}}}
\def\gsim{\raise0.3ex\hbox{$\;>$\kern-0.75em\raise-1.1ex\hbox{$\sim\;$}}}

\def\theta{\vartheta}

\begin{document}

\title{UHE neutrinos from Pop III stars: concept and constraints} 

\author{V.~Berezinsky}
\affiliation{INFN, Laboratori Nazionali del Gran Sasso, I--67010
 Assergi (AQ), Italy}

\author{P.~Blasi}
\affiliation{INAF, Osservatorio Astrofisico di Arcetri, 
Largo E. Fermi 5, 50125 Firenze, Italy}
\affiliation{INFN, Laboratori Nazionali del Gran Sasso, I--67010
 Assergi (AQ), Italy}

\date{\today}

\begin{abstract}
In this paper we reconsider the model of neutrino production during the 'bright phase', first suggested in 1977, in the light of modern understanding of both the role of Pop III stars and of acceleration of particles in supernova shocks. We concentrate on the production of cosmogenic ultra-high energy neutrinos in supernova (SN) explosions that accompany the death of massive Population III stars. Protons are assumed to be accelerated at such SN shocks and produce neutrinos in collisions with CMB photons. In the calculations we deliberately use simplified assumptions which make the physical results transparent. Pop III stars, either directly or through their SN explosions, are assumed to be responsible for the reionization of the universe as observed by WMAP. Since the evolution of massive Pop III stars is much faster than the Hubble rate $H$, we consider the burst of UHE proton production to occur at fixed redshift ($z_b=10$ and $z_b=20$).  We discuss in some detail the problems involved in the formation of collisionless shocks in the early universe as well as the acceleration of charged particles in a medium that has potentially very low pre-existing magnetization, if any at all. The composition of the accelerated particles in Pop III stars explosions is expected to be proton dominated, based upon the predictions of BBN and the Hydrogen-enhanced stellar-wind from primary Pop III stars. A simple calculation is presented to illustrate the fact that the diffuse neutrinos flux from the bright phase burst is concentrated in a relatively narrow energy interval, centered at $E_\nu^c=7.5 \times 10^{15}(20/z_b)^2$~eV. The $\nu_\mu$ flux may be detectable by IceCube without violating the cascade upper limit and without exceeding the expected energetics of SNe associated with Pop III stars. A possible signature of the neutrino production from Pop III stars may be the detection of resonant neutrino events ($\bar{\nu}_e+e^- \to W^- \to$ hadrons) at energy $E_0=6.3\times 10^{15}$~eV. For the burst at $z_b=20$ and $\bar{\nu}_e$-flux at the cascade upper limit, the number of resonant events in IceCube may be as high as 10 events in 5 years of observations. These events have equal energies, $E=6.3\times 10^{15}$~eV , in the form of e-m cascades. Taking into account the large uncertainties in the existing predictions of cosmogenic neutrino fluxes at $E > 10^{15}$ ~eV, we argue that UHE neutrinos from the first stars might become one of the most reliable hopes for UHE neutrino astronomy. 

\end{abstract}


\maketitle

%%%%%%%%%%%%%%%%%%%%%%%%%%%%%%%%%%%%%%%%%%%%%%%%%%%%%%%%%%%%%%%%%%%%%%
\section{Introduction}%
\label{introduction}%
%%%%%%%%%%%%%%%%%%%%%%%%%%%%%%%%%%%%%%%%%%%%%%%%%%%%%%%%%%%%%%%%%%%%%%
Ultra High Energy (UHE) neutrino astronomy at energies above
$10^{16}-10^{17}$~eV may open a new observational window on the
universe. The most reliable prediction for the diffuse fluxes of these
neutrinos can be achieved for the cosmogenic neutrinos \cite{BZ},
produced in $p\gamma$ collisions of UHE protons with CMB photons (see
\cite{cosmogenic} and \cite{reviews} for recent calculations and
reviews). Such reliability mainly follows from three factors: {\it i)}
The flux of parent protons at redshift $z$ of neutrino production can
be normalized to the observed flux of protons at present (redshift
$z=0$); {\it ii)} The number density and energy spectrum of the target
(CMB) photons are both known with high accuracy at any cosmological
epoch $z$; {\it iii)} the interactions of UHE protons with CMB photons
occur at center of mass (cm) energy that does not exceed $\sim GeV$,
well accessible to laboratory experiments and hence  well known. 

On the other hand, the numerical predictions of neutrino fluxes at
Earth are affected by huge uncertainties, mainly due to the unknown
evolution with redshift of the source space density and luminosity, 
the generation spectrum ($\propto E^{-\gamma_g}$) and the maximum 
energy ($E_{\max}^{\rm acc}$) of accelerated protons. Because of these 
uncertainties, any upper bound on the flux of cosmogenic neutrinos becomes 
of the highest importance. For cosmogenic neutrinos this limit is not 
provided by the Waxman-Bahcall bound \cite{WB}, because the latter is 
imposed by the parent proton flux at $z=0$, fixed in the 
calculations. Meanwhile the calculated flux can differ by orders of magnitude due to uncertainties 
in $\gamma_g$, $E_{\max}$ and the evolution of sources \cite{nu2010}. 

A more general upper limit, valid in particular for the cosmogenic
neutrino flux is the {\em cascade upper limit} \cite{cascade}. The
production of UHE neutrinos from pion decays is accompanied by the
production of UHE electrons, positrons and photons, which start an e-m
cascade due to the collisions with CMB photons. The HE photons,
remnants of these cascades, contribute to the diffuse gamma-radiation
with gamma rays of energy reaching several hundreds GeV. Recent
measurements of the diffuse gamma-ray flux by the Fermi-LAT telescope
\cite{fermi} was used in Ref. \cite{BGKO} to put a stronger upper limit on
the diffuse neutrino flux, which excluded many models with detectable 
cosmogenic neutrino flux.

The detectability of cosmogenic neutrinos is further disfavoured by
the recent Auger measurements of UHECR mass composition
\cite{Auger-mass}: both $X_{\max}$ and RMS data strongly suggest that
at energies larger than $2\times 10^{18}$~eV the primaries are nuclei
with steadily increasing mass number $A$. In contrast with these data
the HiRes \cite{HiRes-mass} and Telescope Array \cite{TA} mass composition
agrees well with a pure proton composition.  The cosmogenic neutrino flux produced by primary nuclei is considerably lower than in the case of primary protons \cite{nuNucl}. The Auger results on mass composition, together with the Fermi-based neutrino upper limit suggest a rather grim possibility of detection of cosmogenic neutrinos. 

Here we revise a model of UHE neutrino production related to the 
so-called Population III (Pop III) stars \cite{BO,book}. Recently this population of stars attracted much attention as the most plausible sources of ionizing photons in the early universe, at redshift $z \sim 10-20$, as observed by WMAP \cite{WMAP}. 

Originally, one of the main motivations that led to propose the
existence of Pop III stars was the apparent gap between the chemical
composition of the universe as predicted by BBN, mainly light elements
up to Lithium, and the presence of metals needed for the formation of
Population II stars. Pop III stars bridge this gap by 
synthesizing the first metals in the universe and spreading them
through the 
interstellar and intergalactic medium, where they are later
instrumental for the formation of Pop II stars. For this scenario all Pop III stars or a large fraction of them must
finish their evolution by SN explosions, and therefore these stars  
should be very massive, heavier than $50M_{\odot}$.

After recombination at $z_{\rm rec} \approx 1100$, the baryons in the universe are in the form of cold, neutral, atomic hydrogen and light atoms. However, at redshift $z_{\rm reion}=11.0 \pm 1.4$, according to WMAP  \cite{WMAP}, the baryonic gas is observed to be ionized. The reionization occurred due to photons produced by astrophysical sources. The most plausible sources of such photons are indeed hot Pop III stars, which produce ionizing photons either directly \cite{truran} or as a result of Pop III supernova explosions \cite{SN-reion}.

Pop III stars arise from baryons gravitationally pulled into potential
wells 
formed by dark matter (DM) dominated mini-halos. These mini-halos
collapse 
at $z \sim 20 - 30$ and eventually capture baryons which can form
stellar 
objects. Numerical simulations of this collapse show no fragmentation
of baryonic matter and the first stars form with large masses, spread
over a wide range, with a typical value $M \sim 100 M_{\odot}$ (see \cite{bromm} and references therein). 

At masses $140 M_\odot \leq M_* \leq 260 M_\odot$, Pop III stars are
fully disrupted in SN explosion due to $e^+e^-$
pair-instability. These SNe are called Pair-Instability SNe
(PISN). The energy output of PISN in the form of ejecta is in the
range $10^{51} - 10^{53}$~erg, and could be much higher than for Pop
II and Pop I SNe. In the mass range $M < 140 M_\odot$ and $M > 260
M_\odot$, a massive black hole and disc are produced, with ejection of
gas in the form of jets. These stars are good candidates for GRBs, and
UHE neutrinos from Pop III GRBs were recently studied in
\cite{Meszaros}. 
For further information on Pop III stars and SNe see the reviews 
in \cite{popIII-rev}.

In the following we will refer to this epoch of Pop III star formation
as the {\it bright phase}, an expression often used also to refer to a 
period of enhanced quasar activity (about $\sim 20$ quasars are observed  
at $5.7 \lsim z \lsim 6.4$ \cite{quasars}).

The paper is organized as follows: in section \ref{sec:accel} we discuss some subtle issues related to the formation of shocks in Pop III supernova explosions and in the particle acceleration at such shocks. In section \ref{sec:model} the basic features of the bright-phase model are described. In section \ref{sec:estimates} we present the calculations of the neutrino fluxes and the constraints on such fluxes. The conclusions are presented in section \ref{sec:conclusions}. 

\section{Shock formation and particle acceleration}
\label{sec:accel}

The environment in which Pop III stars explode as SNe is quite unlike that of SNe in the present universe and one may wonder whether particle acceleration may take place at all. The intergalactic medium at redshift $z\sim 10-20$ is unlikely to be magnetized in any appreciable way. This leads to two questions: 1) does a shock develop in the supernova explosions, since such shocks are collisionless, namely mediated by electromagnetic instabilities? 2) if the shock forms, is there enough magnetic turbulence to lead to diffusive particle acceleration up to energies $\sim 10^{19}-10^{20} eV$, needed for neutrino production?  

Numerical simulations of the development of the ion Weibel instability in the absence of a pre-existing magnetic field show that a shock front does in fact form \cite{weibel}. The simulations are carried out for relativistic electron-ion plasmas, while the motion of the plasma ejected from a Pop III star explosion might be newtonian or relativistic depending upon details of the explosion, but the general physical principles that lead to the formation of the shock front should be left unchanged, so we may be confident that a shock wave does indeed form even in a weakly magnetized or unmagnetized medium such as the one that is expected to embed Pop III stars. The fate of the magnetic field behind the shock is not well established: the magnetic filaments which are formed through the Weibel instability might merge to form larger scale magnetic fields, but it is not clear whether this happens before the field may be damped. 

More important than the presence of pre-existing magnetic field is the
reionization of the surrounding gas prior to the SN explosion: a
collisionless shock would hardly develop in a tenuous neutral medium,
such as in the universe soon after recombination. Pop III stars can
naturally fulfill the ionization requirements since, being very hot, 
they are expected to be responsible for the production of intense UV radiation that causes reionization (see e.g.\cite{truran} and references therein).

The issue of particle acceleration is more delicate in that diffusive acceleration at a shock front requires the existence of a turbulent magnetic field upstream and downstream of the shock. The only way that magnetic field can be generated upstream of the shock is through streaming instability, an intrinsically non-linear process (particle acceleration occurs if accelerated particles are able to excite the instability but the instability is excited by accelerated particles). This process is also necessary in SNe in the ISM, and in fact the instability must be driven to its non-linear regime if to reach cosmic ray energies as high as the knee \cite{Emax}. The streaming instability can proceed in a resonant \cite{bell78} or non-resonant \cite{bell04} way, but these are just two sides of the same physical phenomenon \cite{kinetic}. The non-resonant branch grows faster for higher shock velocities \cite{kinetic}.

Once the instability turns non-linear it is very difficult to predict its saturation, which might be dominated by intrinsic dynamical scales (such as the advection time of a fluid element through the shock) or by damping. The issue of damping might be of particular importance in the clouds where Pop III stars originate since the neutral fraction might be appreciable and ion-neutral damping could well shut the instability off. In the absence of damping and assuming a naive extrapolation of quasi-linear theory to the non-linear regime one can estimate the saturation levels of the instability as:
\begin{equation}
\frac{\delta B_{res}^{2}}{8 \pi} = \frac{1}{M_{A}} \rho V_{s}^{2} \xi_{CR}
\end{equation}
in the resonant case, and
\begin{equation}
\frac{\delta B_{nr}^{2}}{4 \pi} = \frac{1}{2} \rho V_{s}^{2}\frac{V_{s}}{c} \xi_{CR}
\end{equation}
in the non resonant case. Here $M_{A}=V_{s}/v_{a}$ is the Alfvenic Mach number, and $v_{a}=B/(4\pi \rho)^{1/2}$ is the Alfven speed. Clearly neither the Alfvenic Mach number nor the Alfven speed are well defined in the non linear regime, but this ambiguity is the price to pay to write down a simple estimate of the magnetic field generated through streaming of accelerated particles upstream of the shock. In all these formulae we assumed that the shock is non-relativistic. Our estimate for the amplified magnetic field is:
\beq
\delta B_{res} \approx 96 B_{\mu}^{1/2} n_{1}^{1/4} V_{9}^{1/2} \xi_{CR}^{1/2} ~ \mu G,
\eeq
in the resonant case and
\beq
\delta B_{nr} \approx 590 n_{1}^{1/2} V_{9}^{3/2} \xi_{CR}^{1/2} ~ \mu G,
\eeq
in the non resonant case. Here $V_{9}=V_{s}/10^{9}\rm cm/s$ and $n_{1}=n/1\rm cm^{-3}$. $\xi_{CR}$ is the fraction of the inflow ram pressure $m_{p}nV_{s}^{2}$ that gets converted to accelerated particles. 

Assuming Bohm diffusion, the acceleration time in the two cases reads
\beq
\tau_{res} (E) \approx 3.5\times 10^{-7} E(eV) B_{\mu}^{-1/2}n_{1}^{-1/4}V_{9}^{-5/2} \xi_{CR}^{-1/2}~\rm s
\eeq
in the resonant case, and 
\beq
\tau_{nr} (E) \approx 5\times 10^{-8} E(eV) V_{9}^{-7/2} n_{1}^{-1/2} \xi_{CR}^{-1/2}~\rm s
\eeq
in the non-resonant case. 

It is important to keep in mind that in the non-resonant case the self-generated turbulence is on spatial scales that are much smaller than the gyration radius of the particles, therefore these modes are not very effective in scattering the particles, despite the large growth rate. In this sense, the assumption of Bohm diffusion for non-resonant modes is poorly justified, and would require an efficient cascade of the modes to much larger scales, an inverse cascade. 

There are at least three different ways to define the maximum energy
of accelerated particles: 1) Time limitation: $\tau(E_{\max}^{\rm acc})\sim T_{ST}$, where $T_{ST}$ is the time when the remnant
enters the Sedov-Taylor phase; 2) Space limitation of the precursor:
$D(E_{max}^{\rm acc})/V_{s}\sim \eta R_{ST}$, where $R_{ST}$ is the
radius of 
the supernova remnant at the beginning of the Sedov phase, and 
$\eta \lesssim 1$; 3) Space limitation on the gyration radius: 
$r_{L}(E_{\max}^{\rm acc})\sim R_{ST}$. The three criteria lead to
different estimates for the value of $E_{\max}^{\rm acc}$.

The Sedov time and radius are related to the ejected mass and to the total energetics through:
\beq
T_{ST} = 70 \frac{M_{ej,\odot}^{5/6}}{\epsilon_{51}^{1/2} n_{1}^{1/3}} ~\rm years
\eeq
and
\beq
R_{ST} = 6.6 \times 10^{18} \left( \frac{M_{ej,\odot}}{n_{1}}\right)^{1/3} \rm cm,
\eeq
where $\epsilon_{51}$ is the total energetics in units of $10^{51}$ erg and $M_{ej,\odot}$ is the mass of the ejecta in units of solar masses.

Among the three criteria for $E_{\max}^{\rm acc}$ listed above, the last one leads to the highest estimated value:
\beq
E_{\max}^{\rm acc} \approx 2\times 10^{17} B_{\mu}^{1/2} n_{1}^{-1/12} V_{9}^{1/2} \xi_{CR}^{1/2} M_{ej,\odot}^{1/3}~\rm eV
\eeq
for the resonant case, and 
\beq
E_{\max}^{\rm acc} \approx 10^{18} n_{1}^{1/6} V_{9}^{3/2} \xi_{CR}^{1/2} M_{ej,\odot}^{1/3}~\rm eV
\eeq
for the non resonant case. One can see that maximum energies as high
as $(2-5)\times 10^{19}$ eV might be reached if the values of the
parameters are pushed to their extremes. 

In principle larger values of $E_{\max}^{\rm acc}$ can be reached if the Pop III 
stars explode in a dense stellar region, so that accelerated particles may 
feel the repeated action of multiple shocks. Such situation occurs   
at high redshifts where the physical volume which hosts the fixed number 
$N$ of pregalactic Pop III stars is $(1+z)^3$ times smaller than at present. Even  
a stronger case might be realized in mini-cluster models, when the Pop III stars
are produced in the relatively small volume of a mini-cluster. In this case the 
spatial scale to be used to determine the maximum energy is the size of the 
region where the shocks emerge, provided the acceleration time remains 
smaller than the loss time scale as plotted in Fig. \ref{fig1}.
Acceleration at multiple shocks has been first discussed in \cite{bell78} 
and is known to result in a hard spectrum: in the asymptotic case of a 
very large number of shocks, the spectrum of accelerated particles tends 
to $\propto E^{-3/2}$.

Larger values of $E_{\max}^{\rm acc}$, up to $10^{20}$~eV, were inferred
in previous literature for the case of acceleration at relativistic shocks with large
Lorentz factors $\Gamma\gg 1$, see e.g. \cite{achterberg}. Usually 
this case is considered for jets, and in particular for GRBs, which 
are often assumed to be potential sources of particles with maximum
energies $\sim 10^{20}$~eV \cite{wax,vietri}. These energies are the consequence of 
the relativistic motion of the outflow in jets, which in principle can shorten the acceleration 
time by a factor $1/\Gamma$. 

Particle acceleration at relativistic shocks requires the presence of strong turbulence generated in the downstream region, in order to avoid particle trapping, typical of relativistic shocks: particle acceleration is possible only for quasi-parallel shocks namely for magnetic fields oriented within an angle $\sim 1/\Gamma$ from the normal to the shock surface. At perpendicular shocks the return probability from downstream tends to vanish thereby making spectra steeper. In fact, even the compression of the large scales in the upstream turbulent fields at the shock surface may lead to spectra steeper than the canonical $E^{-2.3}$ \cite{lemoine}, that is expected for parallel relativistic shocks in the regime of small pitch angle scattering. The presence of strong turbulence may alleviate this problem, so to make GRBs from Pop III stars potential sources of UHECRs with energy $E_{\max}^{\rm acc}\sim 10^{20}$ eV.

The issue of whether particle acceleration can occur in the first stars is also of relevance for the origin of cosmic magnetic fields: in a recent paper \cite{miniati} the authors proposed that if supernovae arising from the death of primeval stars accelerate CRs effectively (though not necessarily to very high energies), the instability induced by the escape of these cosmic rays into the intergalactic medium may lead to the formation of magnetic seeds that can possibly be reprocessed and amplified at later cosmic epochs. 

%%%%%%%%%%%%%%%%%%%%%%%%%%%%%%%%%%%%%%%%%%%%%%%%%%%%%%%%%%%%%%%%%%%%

\section{The model}
\label{sec:model}
%%%%%%%%%%%%%%%%%%%%%%%%%%%%%%%%%%%%%%%%%%%%%%%%%%%%%%%%%%%%%%%%%%%%

Here we consider a model in which the bright phase is powered by SN 
explosions of Pop III stars. 

\begin{figure}[!ht]
\begin{center}
\includegraphics[width=0.95\columnwidth]{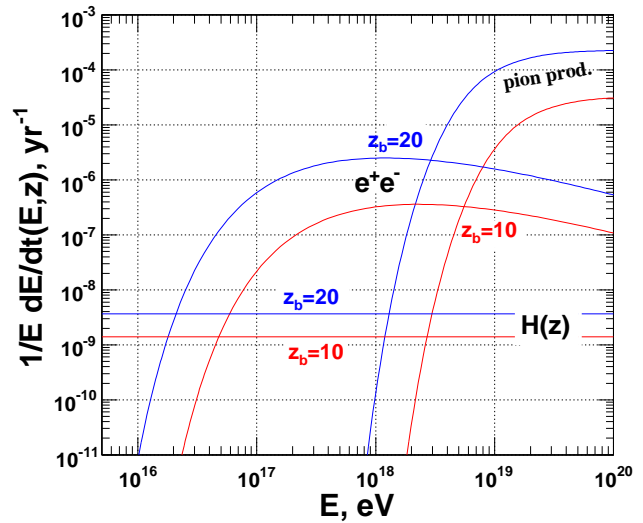}
\caption{Energy losses of protons at cosmological epochs $z_b=10$
  (shown in red) and $z_b=20$ (shown in blue). Adiabatic energy losses
  are given by $H(z)$. Also shown are losses due to pair
  production on CMB, $p+\gamma_{\rm cmb} \to p+e^-+e^+$, and pion
  photo-production, $p+\gamma_{\rm cmb} \to N+$pions. Photo-production
  responsible for the generation of neutrinos dominates at energy above
  the crossing of the pair-production and pion production curves
  ($E_c^b=3\times 10^{18}$~eV at $z_b=20$ and $E_c^b=5.5\times 10^{18}$~eV
  at $z_b=10$). The intersection of the pair-production curve with
  $H(z)$ (adiabatic losses) determines the energy $\varepsilon_{\rm
    pair}$, above which the pair-production energy losses dominate
  ($\varepsilon_{\rm pair}=2.1\times 10^{16}$~eV at $z_b=20$ and
  $\varepsilon_{\rm pair}=5\times 10^{16}$~eV at $z_b=10$).
}
\label{fig1}
\end{center}
\end{figure}
In order to enrich space with metals, the rate of SN explosions must be high and thus Pop III stars have to be massive, typically with 
$M > 50 M_{\odot}$.
Stars in the interval of masses $(140 - 260)M_{\odot}$ undergo
$e^+e^-$ instability and end their evolution with full destruction in
SN explosions. Outside this interval, and most notably at $M > 50
M_{\odot}$, a SN explosion leaves behind a massive 
black hole. In both cases a SN explosion results in the formation of a
shock front where particles from the interstellar medium can be 
accelerated (see section \ref{sec:accel} for the detailed discussion). 
The formation of such (collisionless) shock is expected to be made 
possible because of the ionization of the circumstellar region caused 
by photons produced by either the Pop III star itself or its supernova. 
The chemical composition of the accelerated particles should reflect 
that of the circumstellar medium, which is quite different from 
that in our Galaxy. The primordial gas after BBN is characterised by 
75\% of Hydrogen and by absence of the heavy elements (metals). 
Some heavy element pollution is expected to be caused by multiple 
SN explosions in the same spatial regions.  On the other hand, 
the metallicity of the gas in Pop II star formation regions is expected 
to be relatively low, therefore it appears reasonable to assume that 
the medium around Pop III stars is dominated by light elements. We 
assume here that most of the accelerated particles are protons.

As discussed above, the SN explosion of Pop III stars can result in the 
formation of either non-relativistic or relativistic shock fronts, the 
latter usually associated to the formation of jets. The spectrum of 
accelerated particles in the two cases is usually taken to be 
$\propto E^{-\gamma_g}$ with  $\gamma_g= 2.0$ for non-relativistic shocks 
and $\gamma_g=2.2 - 2.3$ for relativistic shocks, although numerous 
physical effects can change this naive expectation. In both cases 
we assume maximum energy of acceleration $E_{\max}^{\rm acc} \sim 10^{20}$~eV and the minimum energy to be as low as 
$E_{\min} \sim m_{p} c^{2} \sim 10^9$~eV. Increasing $E_{\min}$
results in larger neutrino flux in the calculations below. 

As far as $E_{\max}$ is concerned, the energy  $10^{20}$~eV 
may be problematic in non-relativistic shocks, although particle 
reacceleration at multiple SN shocks may hopefully provide this 
energy (see section \ref{sec:accel}). The situation with  $E_{\max} \sim 10^{20}$~eV is more promising but
still poorly understood in the case of relativistic shocks.

Rather than achieving a quantitatively accurate prediction of the
neutrino 
flux from Pop III stars, our aim here is to develop a simple
calculation 
that catches the main physical ingredients and illustrates in a clear, 
transparent way, the relevant aspects of the problem. The distinctive 
feature of our model is the assumption that the duration of the bright 
phase, occurring at redshift $z_{b}$, is very short compared with the 
Hubble time $H^{-1}(z_b)$ at the same epoch. The latter can be
calculated as 
$H(z_b)=H_0\sqrt{\Omega_m(1+z_b)^3+\Omega_\Lambda}$ 
with $H_0=72$~km/s Mpc,~ $\Omega_m=0.27$, and $\Omega_\Lambda=0.73$. 
At $z_b=10$ the Hubble time is $H(z_b)^{-1} \approx 8\times 10^8$~yr, 
to be compared with the lifetime $\sim 10^{6}$~yr  of a Pop III star 
with $M \sim 100 M_\odot$ on the Main Sequence  (the longest time of 
this star evolution), with the time of particle acceleration, 
$t_{\rm acc} \sim 10^4 - 10^5$~yr, and with the characteristic time 
of photopion energy losses, $\sim 3\times 10^5$~yr, responsible for 
neutrino production (see Fig.~\ref{fig1} for comparison of these times 
at various energies). These estimates serve as a clear justification 
for the assumption of bursting bright phase: {\em all processes are 
assumed to occur at the fixed redshift $z_b$, for which further on we  
consider $z_b=10$ and $z_b=20$ as our benchmark cases (see Fig.~\ref{fig1})}. 
It is interesting to note that models with two bursts of reionization can be formulated \cite{Cen}, as based on taking into account the important role of neutral $H_2$ molecules in star formation. The redshifts of the predicted bursts are $z_1 \approx 11 - 15$ and $z_2 \approx 6$, close to the value of $z_b$ that we consider.

Having in mind the same goal of simplicity of formulae and transparency 
of the calculations, we assume power law spectra with slope 
$\gamma_g = 2.0$ , which does not lead to dramatic differences
compared with similar spectra with $\gamma_{g}\sim 2.2-2.3$. 

%%%%%%%%%%%%%%%%%%%%%%%%%%%%%%%%%%%%%%%%%%%%%%%%%%%%%%%%%%%%%%%%%%%%%%
\section{Assumptions and estimates}%
\label{sec:estimates}
%%%%%%%%%%%%%%%%%%%%%%%%%%%%%%%%%%%%%%%%%%%%%%%%%%%%%%%%%%%%%%%%%%%%%%

The burst of accelerated protons produced during the bright phase at epoch $z_{b}$ produces a spectrum that can be expressed as a function of the energy density $\omega_p(z_b)$ of the same protons at that epoch as:
\beq
n_p(E_b,z_b)= \frac{\omega_p(z_b)}{\ln(E_b^{\max}/E_b^{\min})} E_b^{-2}.  
\label{n(z_b)}
\eeq
Henceforth $\omega_p(z_b)$ will be used at the main normalization parameter of our calculations.

The cosmogenic neutrinos are produced in collisions of UHE protons
with CMB photons through generation of pions, $p+\gamma_{\rm cmb} \to
\pi^{\pm}$+anything. On average a proton transfers $20\%$ of its
energy to the leading pion, and this energy is approximately equally
divided among the four leptons (three neutrinos and one electron), so
that $E_{\nu}= (1/20)E_p$. Photopion production dominates starting at
the proton energy $E_c^b$, where energy losses due to photopion production exceed pair-production losses ($p+\gamma_{\rm cmb} \to p+e^{-}+e^+$). At $z=0$ this condition occurs at proton energy $E_c \approx 6\times 10^{19}$~eV, therefore at epoch $z_b$ this energy is $E_c^b(z_b)=E_c/(1+z_b)$ (see Fig.~\ref{fig1}), since the energy of CMB photon is $(1+z_b)$ times higher. Hence the characteristic neutrino energy in the observed spectrum is 
\beq
E_\nu^c=\frac{1}{20}\frac{E_c}{(1+z_b)^2}=7.5\times 10^{15}\left (
\frac{20}{1+z_b}\right )^2~
{\rm eV},
\label{Enu-min}
\eeq
where an extra factor $(1+z_b)$ appears because of the redshifted
neutrino energy.
Below $E_\nu^c$, the neutrino spectrum becomes flatter, being produced by 
pions moving backwards in the cm-system. Above $E_\nu^c$, the neutrino 
spectrum follows the parent proton spectrum, $\propto E^{-2}$. 
Eq.~(\ref{Enu-min}) thus gives the neutrino energy where detectability 
reaches its maximum. The main contribution to the number of detected
neutrinos is given by some energy interval centered at  $E_\nu^c$.

For the calculation of the neutrino spectrum it is convenient, 
following \cite{BZ,BG88,BGG-prd}, to introduce the so-called 
'unmodified' proton spectrum, calculated taking into account 
only adiabatic energy losses due to redshift. This proton spectrum 
at $z=0$ is  
\beq 
n_p^{\rm unm}(E)= \frac{\omega_p(z_b)}{(1+z_b)^4}
\frac{1}{\ln(E_{\max}/E_{\min})}E^{-2}.
\label{n_unm}
\eeq
Eq.(\ref{n_unm}) is easy to understand because $\omega_p(z_b)/(1+z_b)^4$ 
is the unmodified energy density of protons at $z=0$: 
$(1+z_b)^3$ is due to the expansion of the universe and the 
extra factor $(1+z_b)$ accounts for adiabatic energy losses. 

The spatial density of UHE neutrinos at $z=0$ can now be readily 
calculated using the unmodified proton spectrum and taking into 
account that each proton in one photonuclear $p\gamma$ collision 
produces three neutrinos and the energy of a neutrino $E_\nu$ is 
connected with the energy $E$ of the 'unmodified' proton as 
$E_{\nu}=(1/20)E$, independent of the redshift $z$ when the collision 
occurs, because adiabatic energy losses of neutrinos and UHE protons 
are identical. 

Thus, one obtains at $E_\nu \geq E_\nu^c$ : 
\beq
n_{\nu}(E_{\nu})dE_{\nu}=\frac{3f_{\pi^\pm}}{20}\frac{\omega_p(z_b)}
{(1+z_b)^4}\frac{1}{\ln(E_{\max}/E_{\min})}\frac{dE_{\nu}}{E_{\nu}^2},
\label{n_nu}
\eeq
where $f_{\pi^\pm} \approx 2/3$ is the fraction of charged pions.

The maximum neutrino energy at $z=0$ can be estimated as
\beq 
E_\nu^{\max}=\frac{1}{20} E_{\max}=\frac{1}{20} \frac{E_b^{\max}}{1+z_b}.
\label{Enu-max}
\eeq 

The neutrino flux, 
\beq
E^2_{\nu} J_{\nu}(E_\nu)=0.1\frac{c}{4\pi}\frac{\omega_p(z_b)}{(1+z_b)^4}
\frac{1}{\ln (E_{\max}/E_{\min})},
\label{nu-flux}
\eeq
is fully determined by the value of the basic parameter 
$\omega_p(z_b)/(1+z_b)^4$. 

We can normalize this parameter to the future IceCube sensitivity 
\cite{IceCube} $E^2J_{{\nu}_{\mu}}(E)=3\times 10^{-9}$~GeV/cm$^2$ s sr. (The 
present upper limit of IceCube is a factor of 3 higher). From this 
condition the basic parameter must be 
\beq
\omega_p(z_b)/(1+z_b)^4 \geq 9.5 \times 10^{-7}~{\rm eV~cm}^{-3}.
\label{omega-num}
\eeq

We can estimate now the maximum energy of acceleration 
$E_{\max}^{\rm acc}=E_b^{\max}$ discussed in section \ref{sec:accel}. 
The rigorous limit follows from the condition that pion production 
dominates upon pair production (see Fig.~\ref{fig1}):
\beq
E_b^{\max}=E_c(z_b)=\frac{E_c}{1+z_b}=3\times 10^{18} \left( \frac{20}{1+z_b}  
\right)~{\rm eV}.
\label{Emax-rig}
\eeq
The same limit follows from the condition which provides an appreciable neutrino flux, $E_\nu^{\max} \geq E_\nu^c$. In the present estimates we assume that $E_b^{\max} \sim 10^{20}$~eV. The case $E_b^{\max} \gsim 1\times 10^{19}$~eV needs more detailed calculation of the neutrino spectra and will be presented in a forthcoming paper.

In the following we discuss the constraints on the cosmogenic neutrino 
flux from the bright phase. 

The production of UHE neutrinos from pion decays is accompanied 
by e-m cascade radiation, which provides us with an upper limit on 
the neutrino flux \cite{cascade}. The neutrino flux given by 
Eq.~(\ref{nu-flux}) with energy density (\ref{omega-num}) must respect 
this limit as given by the recent Fermi observations \cite{fermi}. 
In \cite{BGKO} this limit is expressed in terms of an upper limit on 
the energy density of the cascade radiation 
$\omega_{\rm cas}^{\max} \leq 5.8\times 10^{-7}$~eV/cm$^3$. We can estimate 
the cascade energy density in our calculations from $\omega_p(z_b)$ 
evaluated in Eq.~(\ref{omega-num}). 

At energies $E_b \geq \varepsilon_{\rm pair}$ (see Fig.\ref{fig1}), 
protons lose energy on time scales shorter than the Hubble time 
$H^{-1}(z_b)$, producing $e^+e^-$ pairs and pions. The energy density $\omega_{\rm cas}^b$ of 
the cascade radiation at epoch $z_b$ can be written as  
\beq
\omega_{\rm cas}^b \lsim \int_{\varepsilon_{\rm pair}}^{E_b^{\max}} 
n_p(E_b) E_b dE_b ,
\label{omega-cas_b}
\eeq
and for the cascade energy density at $z=0$ we have
\ba
\label{eq:omega_cas}
\omega_{\rm cas} \lsim \frac{\omega_p(z_b)}{(1+z_b)^4}
\left [ \ln \frac{E_b^{\max}}{\varepsilon_{\rm pair}}\right ]
\left [\ln\frac{E_b^{\max}}{E_b^{\min}} \right]^{-1}\\  \nonumber 
= 2.9\times 10^{-7}{\rm eV/cm}^3. 
\ea
One can see that the cascade energy density (\ref{eq:omega_cas}), which 
follows from neutrino production in the bright phase does not exceed 
the upper limit $\omega_{\rm cas}^{\max}=5.8\times 10^{-7}$~eV/cm$^3$, obtained in 
\cite{BGKO} from Fermi observations. In fact, this upper limit is derived 
for the cascade spectrum produced at redshift $z=0$, while in our case 
the cascade is produced at large $z_b$ and its spectrum is
redshifted. This makes the upper limit higher than $\omega_{\rm cas}^{\max}$ from \cite{BGKO}. 

Finally, we address the issue of whether the energy density of protons 
in Eq.~(\ref{omega-num}) which is needed to warrant detectability of 
the neutrino flux in IceCube can be provided by Pop III SN explosions. 
For this, we estimate a fraction $\xi$ of cosmological baryon mass 
processed by Pop III pre-supernovae. 

From the energy density of UHE protons $\omega_p(z_b)=W_{\rm SN}^p n_*(z_b)$, 
provided by the SN energy release in the form of UHE protons $W_{\rm SN}^p$ 
and from the space density of Pop III presupernovae $n_*$, one obtains 
\beq
\xi = \frac{\omega_p(z_b)}{(1+z_b)^4}\frac{M_* (1+z_b)}
{W_{\rm SN}^p \Omega_b \rho_{\rm cr}}.
\label{xi}
\eeq
Here we used the equality $M_*n_*(z_b)=\xi \rho_b(z_b)$, where 
$M_* \sim 100 M_\odot$ is a typical mass of Pop III presupernova and 
$\rho_b(z_b)=\Omega_b\rho_{\rm cr}(1+z_b)^3$ is the total baryonic mass 
density at epoch $z_b$, where for the baryonic gas density at $z=0$ we 
use the WMAP \cite{WMAP} value 
$\Omega_b\rho_{\rm cr} = 4.27\times 10^{-31}$~g/cm$^3$. 
For $M_*=100 M_\odot$ the SN energy release can be conservatively 
estimated as $W_{\rm SN}^p \sim 10^{51}$~erg. As a result we obtain 
for the case of Eq.~(\ref{omega-num}) the reasonable value 
$\xi=1.4\times 10^{-2}$, i.e. about $1\%$ of the total baryonic mass 
processed by Pop III presupernovae.  

If to assume that each SN explosion leaves behind a black hole with mass 
$M_{\rm bh} \sim 3 M_{\odot}$, the fraction of baryonic mass in the form of 
these black holes is only $\eta = \xi M_{\rm bh}/M_* \sim 4 \times 10^{-4}$.
%%%%%%%%%%%%%%%%%%%%%%%%%%%%%%%%%%%%%%%%%%%%%%%%%%%%%%%%%%%%%%%%%

\subsection{$\bar{\nu}_e+e^- \to W^- \to hadrons~ resonance.$}
\label{sec:resonance}
%%%%%%%%%%%%%%%%%%%%%%%%%%%%%%%%%%%%%%%%%%%%%%%%%%%%%%%%%%%%%%%%%%%%%%
The resonant production of hadrons by UHE $\bar{\nu}_e$-neutrinos is
a remarkable feature and signature of the Pop III bright phase model. 

The resonant production of W-bosons was first proposed by S.L. Glashow 
in 1960 \cite{glashow} for the generation of high energy muons in the 
reaction $\bar{\nu}_e+e^- \to W^- \to \mu^- + \bar{\nu}_\mu$. 
In 1977 Berezinsky and Gazizov \cite{BG77} suggested another channel 
of the Glashow resonance $\bar{\nu}_e+e^- \to W^- \to $~hadrons for 
the detection of UHE neutrinos. The resonant neutrino energy is given by 
\beq 
E_0=m_W^2/(2m_e)=6.3\times 10^6~{\rm GeV},
\label{Eres}
\eeq
and the rate of resonant events $\nu_{\rm res}$ in a detector with total number of electrons $N_e$ is determined by the exact formula derived in \cite{BG77}: 
\beq
\nu_{\rm res}=2\pi N_e \sigma_{\rm eff} J_{\bar{\nu}_e}(E_0) E_0 ,
\label{nu-res}
\eeq
where $J_{\bar{\nu}_e}(E_0)$ is the diffuse flux at the resonant energy 
$E_0$, $N_e=(5/9)M/m_H$ is the number of electrons in the underground 
detector with mass $M$, $2\pi$ is the solid angle at which a 
deep-underground detector is open for resonant neutrinos (for the 
detailed calculations see \cite{BG81}), 
$\sigma_{\rm eff}=(3\pi /\sqrt{2})G_F=3.0\times 10^{-32}$~cm$^2$ is 
the effective cross-section obtained by integration of the 
Breit-Wigner cross-section over energy. 

The  flux of $\bar{\nu}_e$-neutrinos produced in $p\gamma_{\rm cmb}$ 
collisions is strongly suppressed and appears mostly due to oscillations 
as has been already anticipated in \cite{BG77}. According to recent 
calculations, $J_{\bar{\nu}_e}(E)$ is approximately the same for all 
flavours, namely it is about $1/6$ of the all-flavour neutrino flux.  

For the application of the Glashow resonance to the Pop III bright phase, 
we have to first determine $z_b$. It can be found from the condition that 
the energy of the parent proton for the resonant neutrino at $z_b$ must 
be above the critical energy $E_c/(1+z_b)$, at which the energy losses 
for pion and $e^+e^-$-pair production are equal. Using $E_c = 6\times 10^{19}$~eV one has  
\beq
1+z_b=\left (\frac{E_c}{20 E_0} \right )^{1/2} = 21.8,   
\label{z_b-glashow}
\eeq
which coincides with our assumption $z_b=20$.

The attractive feature of this model is that it imposes milder constraints 
on $E_{\max}^{\rm acc}$. Indeed, the energy of the parent proton 
for resonant neutrino at $z_b$ is $E_p(z_b)=20E_0(1+z_b)=2.8\times 10^{18}$~eV, 
which is considerably lower than $E_{\max}^{\rm acc}$ discussed in 
section \ref{sec:accel}. Below  we assume $E_{\max}^{\rm acc}= 1\times 10^{19}$~eV.

We calculate now the rate of the resonant events in IceCube with
effective mass $M \sim 1\times 10^9$~t. According to Eq.~(\ref{nu-res}), 
there is only one unknown quantity for the calculations, the flux 
$J_{\bar{\nu}_e}(E_0)$, which we take as maximal, namely corresponding
to the cascade upper limit $\omega_{\rm cas}^{\max}=5.8\times 10^{-7}$~eV/cm$^3$. 
Using Eq.~(\ref{nu-flux}) for the neutrino flux with 
$\omega_p(z_b)/(1+z_b)^4$ from Eq.~(\ref{eq:omega_cas}), and assuming 
equipartition of neutrino flavours with a fraction $1/6$ for 
$\bar{\nu}_e$, we obtain the rate of the resonant events in IceCube as 
\beq
\nu_{\rm res}=\frac{0.1 c}{12} \frac{N_e \sigma_{\rm eff}}
{\ln (E_b^{\max}/\varepsilon_{\rm pair})} 
\frac{\omega_{\rm cas}^{\max}}{E_0} 
%= 1.4~{\rm yr}^{-1} .
\label{nu_res1}
\eeq
In more accurate calculations, where one takes into account that 
decaying pions transfer only half of their energy into e-m cascade,
$E_b^{\max}$ in Eq.~(\ref{nu_res1}) should be substituted by 
$\sqrt{E_c^bE_b^{\max}}$, and the rate of events
reaches $\nu_{\rm res}= 1.3~{\rm yr}^{-1}$.  In fact, this rate is 
not the maximal one, because as already mentioned above, 
$\omega_{\rm cas}^{\max}=5.8\times 10^{-7}$~eV/cm$^3$ was obtained for 
$z=0$, while for $z_b=20$ this limit is higher. Most probably each 
resonant event with tremendous energy release $6.3\times 10^{15}$~ eV 
in the form of nuclear and electromagnetic cascades will be detected 
in IceCube. Ten well identified events with equal energies during five 
years of observations is a good enough signature of Pop III bright 
phase model with $z_b \geq 20$.

%%%%%%%%%%%%%%%%%%%%%%%%%%%%%%%%%%%%%%%%%%%%%%%%%%%%%%%%%%%%%%%%%%
\section{Conclusions}
\label{sec:conclusions}
%%%%%%%%%%%%%%%%%%%%%%%%%%%%%%%%%%%%%%%%%%%%%%%%%%%%%%%%%%%%%%%%%

There are theoretical and observational indications that the first star formation, the galaxy formation and AGN went through a phase of enhanced activity at high redshifts, $z \sim 10 - 20$ (e.g. low metallicity stars and the observation of about 20 quasars at $z \sim 5 - 6.4$ \cite{quasars}).  

In this paper we concentrate upon a bright phase in the stellar evolution, in the form of Pop III stars. These astrophysical objects play an important role in at least two ways: 1) they enrich the universe with the metals that appear to be necessary for the formation of Pop II stars, thereby bridging the gap between the chemical composition of the universe as predicted by BBN and the one observed in nowadays universe. 2) Pop III stars may reionize the universe, at a redshift that WMAP observations measure as $z_{\rm reion}=11.0\pm 1.4$

Both the metal enrichment and the reionization require that Pop III stars are short-lived and therefore very massive stars, typically  with $M_* \gsim 100 M_\odot$, which explode as supernovae. In the mass range $140 M_\odot \lsim M_* \lsim 260 M_\odot$ the star is fully disrupted due to the $e^+e^-$ pair-instability. In the mass range outside this interval a massive black hole and disc are produced, with ejection of gas in the form of jets. In both cases a SN explosion results in a shock where particle acceleration may take place. In the low-mass limit $M \sim (15 - 40) M_\odot$ the energy release is $W_{\rm SN} \approx 2 \times 10^{51}$~erg \cite{Joggerst}, at large masses $M > 100 M_\odot$ it may exceed $10^{53}$~erg. In case of a very massive black hole the energy output can be much higher due to the Blandford-Znajek effect.  

Here we concentrated on the possibility that diffusive particle
acceleration may take place at the shocks that are generated when Pop
III stars explode. The medium in which these SN explosions occur is
expected to be made mainly of hydrogen (75 \%), with a small
contamination of helium and other light elements. Although a metal 
contamination must appear and increase once the bright phase starts, 
we expect that this represents a weak effect. Therefore, we 
considered protons as the main component of accelerated particles. 

We discussed at length three delicate aspects of the problem of 
acceleration: 1) the formation of collisionless shocks around 
primordial stars; 2) the generation of magnetic field; 
3) particle scattering and acceleration. 

Pop III stars are the most natural sources of reionization of 
the universe due to their high temperature and luminosity \cite{truran}.
The reionization of the universe is a crucial ingredient of our
discussion, since in the absence of such a phenomenon the collisionless shocks 
asociated with SN explosions would hardly form or would have very peculiar 
characteristics. These shocks are formed because of the mediation of 
electromagnetic instabilities such as the Weibel instability that was 
found to be effective even in the absence of a pre-existing magnetic 
field. The generation of magnetic field is just a different aspect of 
the problem of shock formation, in that the presence of magnetic field 
is the very reason why particle motion is slowed down, thereby fulfilling 
the Rankine-Hugoniot conditions at the shock surface. Magnetic field can 
also be formed downstream of the shock if the upstream plasma contains density inhomogeneities that induce shock corrugation and eventually lead to eddies that may considerably amplify a magnetic seed \cite{giacalone}. For relativistic shocks, ampliÞcation of the magnetic field by a macroscopic turbulent 
dynamo triggered by the Kelvin-Helmholtz shear instability has been investigated in \cite{zhang}.

It is worth stressing that in these cases the magnetic field is formed behind the shock surface and advected downstream. It is therefore not useful
in terms of scattering the particles upstream of the shock, a
necessary condition for particle acceleration. Magnetic field can be 
produced upstream of the shock through streaming instability excited by 
accelerated particles. This instability can be excited in a resonant 
or non-resonant way, the former being in general slower for fast shocks 
but more effective in scattering accelerated particles. We found that 
these processes can hardly allow particles to reach energies in 
excess of $10^{19}$ eV for fast but non-relativistic shocks, although higher $E_{\max}$ can be possibly obtained if numerous SNe 
explode in the same region. 

Higher maximum energies, in excess of $10^{20}$ eV have been widely discussed and claimed to be achievable in the case of relativistic 
motion of the SN ejecta \cite{vietri,achterberg} that are thought to occur in GRB afterglows (see however discussion in section \ref{sec:accel}).

UHE neutrinos in our model are produced in $p\gamma_{\rm cmb}$
collisions. The energy of CMB photons is $(1+z_b)$ times higher than at $z=0$, 
and the density of these photons is $(1+z_{b})^3$ times larger. 
We assume that a burst of UHE proton generation occurs at redshift 
$z_b$ and consider two values $z_b=10$ and $z_b=20$ as benchmark
cases. 
The generation spectrum is assumed to be $\propto E^{-2}$ with 
$E_{\max}^{\rm acc} = 1\times 10^{20}$~eV or somewhat less. 
With these assumptions, the neutrino flux, given by Eq.~\ref{nu-flux}), 
is fully determined by the basic parameter represented by the energy 
density $\omega_p(z_b)$ of UHE protons at epoch $z_b$, recalculated at 
the present epoch $\omega_p= \omega_p(z_b)/(1+z_b)^4$ 
(see Eq.~(\ref{nu-flux}) ). In order to have the muon-neutrino flux 
detectable by IceCube, this parameter must be 
$\omega_p \gsim 1\times 10^{-6}$~eV/cm$^3$. The dominant number of 
detected events is confined within a limited energy interval centered 
at energy $E_\nu^c=7.5\times 10^{15}(20/z_b)^2$~eV (Eq.~(\ref{Enu-min}). 
At this energy the detected flux has a weak maximum. The predicted flux 
with the value of $\omega_p$ given above is detectable by IceCube and 
respects the cascade upper limit on cosmogenic neutrinos and 
SN energetics for Pop III stars.

A unique signature of the Pop III burst model may be provided by 
$\bar{\nu}_e$ neutrinos interacting in a detector through the resonant 
reaction  $\bar{\nu}_e + e^- \to W^- \to$ hadrons. The resonant energy
of the neutrino is given by Eq.~(\ref{Eres}) as $E_0=6.3\times 10^{15}$~eV, 
and the rate of neutrino events in a detector is determined by one 
unknown quantity, the neutrino flux at energy 
$E_{0}$, $J_{\bar{\nu}_e}(E_0)$, see Eq.~({\ref{nu-res}). We take this flux 
as to saturate the cascade upper limit corresponding to the cascade  
energy density $\omega_{\rm cas}^{\max} = 5.8 \times 10^{-7}$~eV/cm$^3$, 
though in the model with fixed $z_b$ this limit is expected to be higher. The 
redshift of the burst $z_b$, or more precisely its lower limit, can be 
found from the condition that the energy $E_p(z_b)$ of a proton, 
parent of the resonant neutrino, must have energy higher than 
$E_c(z_b)$, at which pion energy losses at epoch $z_b$ exceed those 
for $e^+e^-$ production. This condition determines the burst redshift 
as $z_b=20.8$, see Eq.~(\ref{z_b-glashow}). The resonant events can be 
observed in IceCube through the Cerenkov light from  nuclear and e-m 
cascades with energy $E_0$. In IceCube the predicted frequency of resonant 
events for the flux at the cascade upper limit is rather low, about 10 events 
in 5 years, but with a tremendous energy deposit, the same for all events. 
This may provide us with a reasonable signature of the model discussed
here. These events are accompanied by muons with energies of order 
$0.5 E_\nu^c \sim 4\times 10^{15}$~eV. 
%%%%%%%%%%%%%%%%%%%%%%%%%%%%%%%%%%%%%%%%%%%%%%%%%%%%%%%%%%%%%%%%%%%%%%
%\section*{Acknowledgments}%
%%%%%%%%%%%%%%%%%%%%%%%%%%%%%%%%%%%%%%%%%%%%%%%%%%%%%%%%%%%%%%%%%%%%%%

\end{document}